\documentclass[aps,prl,reprint,amsmath,amssymb,showpacs,preprintnumbers,superscriptaddress]{revtex4-1}
\usepackage{CJK}
\usepackage{graphicx}
\usepackage{dcolumn}
\usepackage{bm}
\usepackage{color}
\usepackage{hyperref}

\allowdisplaybreaks

\begin{document}
\begin{CJK*}{UTF8}{}
\title{Dynamical Synthesis of $^4$He in the Scission Phase of Nuclear Fission}
\CJKfamily{gbsn}
\author{Z. X. Ren}
\affiliation{State Key Laboratory of Nuclear Physics and Technology, School of Physics, Peking University, Beijing 100871, China}
\author{D. Vretenar}
\affiliation{Physics Department, Faculty of Science, University of Zagreb, 10000 Zagreb, Croatia}
\affiliation{State Key Laboratory of Nuclear Physics and Technology, School of Physics, Peking University, Beijing 100871, China}
\author{T. Nik\v si\' c}
\affiliation{Physics Department, Faculty of Science, University of Zagreb, 10000 Zagreb, Croatia}
\affiliation{State Key Laboratory of Nuclear Physics and Technology, School of Physics, Peking University, Beijing 100871, China}
 \author{P. W. Zhao}
\affiliation{State Key Laboratory of Nuclear Physics and Technology, School of Physics, Peking University, Beijing 100871, China}
\author{J. Zhao}
\affiliation{Center for Circuits and Systems, Peng Cheng Laboratory, Shenzhen 518055, China}
\author{J. Meng}
\affiliation{State Key Laboratory of Nuclear Physics and Technology, School of Physics, Peking University, Beijing 100871, China}

\begin{abstract}
 In the exothermic process of fission decay, an atomic nucleus splits into two or more independent fragments. Several aspects of nuclear fission are not properly understood, in particular the formation of the neck between the nascent fragments, and the subsequent mechanism of scission into two or more independent fragments. Using an implementation of time-dependent density functional theory, based on a relativistic energy density functional and including pairing correlations, we analyze the final phase of the process of induced fission of $^{240}$Pu, and show that the timescale of neck formation coincides with the assembly of two $\alpha$-like clusters. Because of its much larger binding energy, the dynamical synthesis of $^{4}$He in the neck predominates over other light clusters, e.g., $^{3}$H and $^{6}$He. At the instant of scission the neck ruptures exactly between the two $\alpha$-like clusters, which separate because of the Coulomb repulsion and are eventually absorbed by the two emerging fragments.
 The mechanism of light charged clusters formation at scission could also be linked to ternary fission.
\end{abstract}

\date{\today}

\maketitle

\end{CJK*}


Ever since at the turn of the twentieth century $^{4}$He was identified as the mysterious $\alpha$ particle that can spontaneously be emitted from a heavy nucleus, radioactive decays driven by the strong interaction have remained a fascinating topic of research in nuclear physics.
Not all spontaneous nuclear decay modes have been identified and, in addition to $\alpha$ decay and the fission process in which a nucleus splits into two fragments of comparable masses, the emission of clusters ranging from $^{14}$C to $^{34}$Si has been observed in the actinide region.
Nuclear fission is an exothermic dynamical process, in which a quasistationary initial state evolves in time, spontaneously or following a perturbation, through a sequence of increasingly deformed shapes. The nucleus reaches the outer saddle point on the deformation energy surface, and continues to deform while a neck appears that eventually becomes so thin that scission occurs, and the system separates into independent fragments. A fissioning nucleus generally breaks into two heavy fragments of comparable mass, but fission into three fragments is also dynamically possible. The latter process is termed ternary fission and it was observed and described already in the 1940s \cite{san-tsiang47}.

Once in every few hundred fission events, a light charged cluster is emitted in addition to two heavy fragments. In most cases these are $\alpha$ particles, with a much smaller contribution of $^{3}$H and $^{6}$He nuclei \cite{loveland67,halpern71}. For every hundred $\alpha$-particles emitted in ternary fission, typically less than ten tritons, and less than two $^{6}$He nuclei are observed, while the number of other light fragments, like deuteron, $^{3}$He, or $^{6}$Li, is much smaller. The energy and direction of the light cluster, with respect to the fission axis, is determined by the strong Coulomb field of the heavy fragments. For instance, the nearly Gaussian energy distribution of the $\alpha$ particles lies in the interval between 6 and 30 MeV, with a broad peak at 16 MeV, irrespective of the fissioning species \cite{wag04, ver08}, and the angular distribution is strongly peaked at about $80^\circ$ to $85^\circ$ with respect to the lighter of the two heavy fragments \cite{halpern71}. The energy and angular distributions of all light charged clusters indicate that they are emitted in coincidence with scission \cite{halpern71}.  Therefore, ternary fission presents evidence for the formation of light clusters, predominantly $\alpha$ particles, during the final stage of the fission process \cite{wuenschel14,ropke21,denisov21}.

Fission dynamics has been described using a variety of phenomenological and microscopic approaches \cite{bender20,schmidt18,schunck16}. The two basic microscopic methodologies are the time-dependent generator coordinate method (TDGCM), and the time-dependent density functional theory (TDDFT).
The former provides a fully quantum mechanical description of the fission process, starting from the ground state and followed by an adiabatic evolution in time of collective degrees of freedom all the way to scission and the emergence of fission fragments.
However, fission dynamics from the outer fission barrier (saddle point) to the scission configuration is dissipative rather than adiabatic.
The TDDFT framework, which automatically includes one-body dissipation, is more appropriate for modeling the nonequilibrium saddle-to-scission phase of the fission process \cite{bulgac20}.
TDDFT studies of fission dynamics began with the pioneering work of Negele {\it et al.} \cite{negele78}, and have recently been reintroduced, without the inclusion of dynamical pairing correlations \cite{simenel14,goddard15}, with dynamical pairing in the BCS approximation \cite{scamps15} and the HFB method \cite{bulgac19}, and even with the approximate treatment of one-body fluctuations \cite{tanimura17}.

Generally, the mechanism of neck formation and scission in the fission process is not well understood.
Even the point of scission, where separated fragments emerge at the instant of neck rupture, cannot be uniquely defined.
For instance, in the {\em statistical scission-point} model \cite{lemaitre15,lemaitre19}, that assumes a thermodynamic equilibrium at scission, the system at scission is treated as a microcanonical ensemble where all available configurations are equiprobable.
The {\em random neck rupture} model relies on a sequence of instabilities  \cite{brosa90}, and assumes that the neck snaps at some random points along its length. In microscopic approaches, both geometrical and dynamical definitions of the scission configuration have been considered. Geometrical definitions include the criterion of vanishing density between the fragments, and the expectation value of a neck operator that gives a measure of the number of particles in the neck \cite{han21}. In a dynamical approach, the scission configuration can be defined in terms of the ratio of the nuclear and Coulomb interaction energies in the neck region \cite{bonneau07}, as well as using a quantum localization method based on the partition of orbital wave functions into two sets belonging to the prefragments \cite{younes11}.

In this work, the dynamics of neck formation and nuclear scission is studied within the TDDFT framework. For the details of the particular implementation of TDDFT that we employ to model induced fission, we refer the reader to the Supplemental Material (SM)~\cite{SM} and Refs.~\cite{ren20LCS, ren20O16, ebata10TDBCS, Scamps13TDBCS}.

\begin{figure*}[tbh!]
\centering
\includegraphics[width=0.75\textwidth]{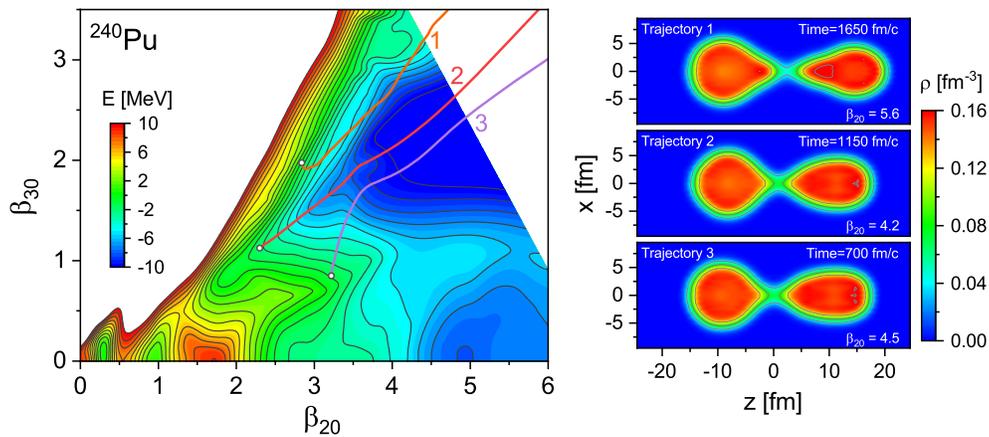}\\
 \caption{Self-consistent deformation energy surface of $^{240}$Pu in the plane of quadrupole-octupole axially symmetric deformation parameters, calculated with the relativistic density functional PC-PK1 functional and a monopole pairing interaction. Contours join points on the surface with the same energy (in MeV). The curves denote the TDDFT fission trajectories for three initial points on the energy surface. For these trajectories the panel on the right displays the corresponding density profiles (color code in fm$^{-3}$) in the $x$-$z$ coordinate plane, at times immediately prior to the scission event.}
 \label{fig:240PES}
\end{figure*}

In the left panel of Fig.~\ref{fig:240PES}, we display the self-consistent deformation energy surface of $^{240}$Pu.
It is calculated with the multidimensionally constrained relativistic mean-field model~\cite{Lu12MDCRMF, Lu14MDCRMF, zhou16MDCRMF} based on the relativistic energy density functional PC-PK1~\cite{zhao10} and BCS approximation with a monopole pairing interaction~\cite{Ring2004manybody} (cf. SM~\cite{SM} for details), and shown as a function of the two collective coordinates: the axial quadrupole ($\beta_{20}$) and octupole ($\beta_{30}$) deformation parameters, that correspond to the nuclear elongation and mass asymmetry, respectively. The equilibrium minimum is located at $\beta_{20}\approx 0.3$ and $\beta_{30}=0$, and it appears slightly soft in the octupole direction. We also note the isomeric minimum at $\beta_{20}\approx 0.9$ and $\beta_{30}=0$, as well as the two fission barriers, and the fission valley at large deformations.

The dots in the left panel of Fig.~\ref{fig:240PES} denote three characteristic initial points on the energy surface for calculation of fission trajectories, where the initial wave functions are obtained in three-dimensional lattice space based on the inverse Hamiltonian and Fourier spectral methods~\cite{ren17dirac3d, ren19LCS}.
Since TDDFT effectively describes the classical evolution of independent nucleons in self-consistent mean-field potentials, this approach cannot be applied to fission dynamics in the classically forbidden region of the collective space \cite{bender20,schunck16,bulgac20,simenel14}.
The initial point for the TDDFT evolution is usually taken below the outer barrier \cite{scamps18,bulgac19}, and the three points shown in the left panel of Fig.~\ref{fig:240PES} correspond to energies approximately 1 MeV below the equilibrium minimum.
Given the initial single-nucleon quasiparticle wave functions and occupation probabilities, TDDFT models a single fission events by propagating the nucleons independently toward scission and beyond.
At each step in time the single-nucleon Hamiltonian is determined from the time-dependent densities, currents, and pairing tensor (cf. SM~\cite{SM} for details), and, therefore, the time evolution includes the one-body dissipation mechanism.

For the three fission trajectories, the panel on the right of Fig.~\ref{fig:240PES} displays the corresponding isodensities (in units of fm$^{-3}$) in the $x$-$z$ coordinate plane, at times immediately prior to the scission event. Even though the lengths of the fission trajectories in the collective space of deformation parameters are not dramatically different, the time it takes to reach the scission configuration varies from 1650 fm/c (trajectory 1), to 1150 fm/c (trajectory 2) and, finally, 700 fm/c (trajectory 3). The large differences in time can be attributed to the self-consistent potentials in which the system evolves toward scission along the three trajectories and to dynamical pairing correlations \cite{bulgac19}. These times should be compared with the average timescale for the evolution of a nucleus from the compound system in equilibrium to the formation of fragments: $(6-15)\times10^3$~fm/c \cite{hinde92}. The isodensities in Fig.~\ref{fig:240PES} exhibit a typical mass asymmetry of the nascent fragments, and also the low-density neck region  characterized by the competition between the repulsive long-range Coulomb interaction and the short-range nuclear attraction.

\begin{figure}[tbh!]
\centering
\includegraphics[width=0.45\textwidth]{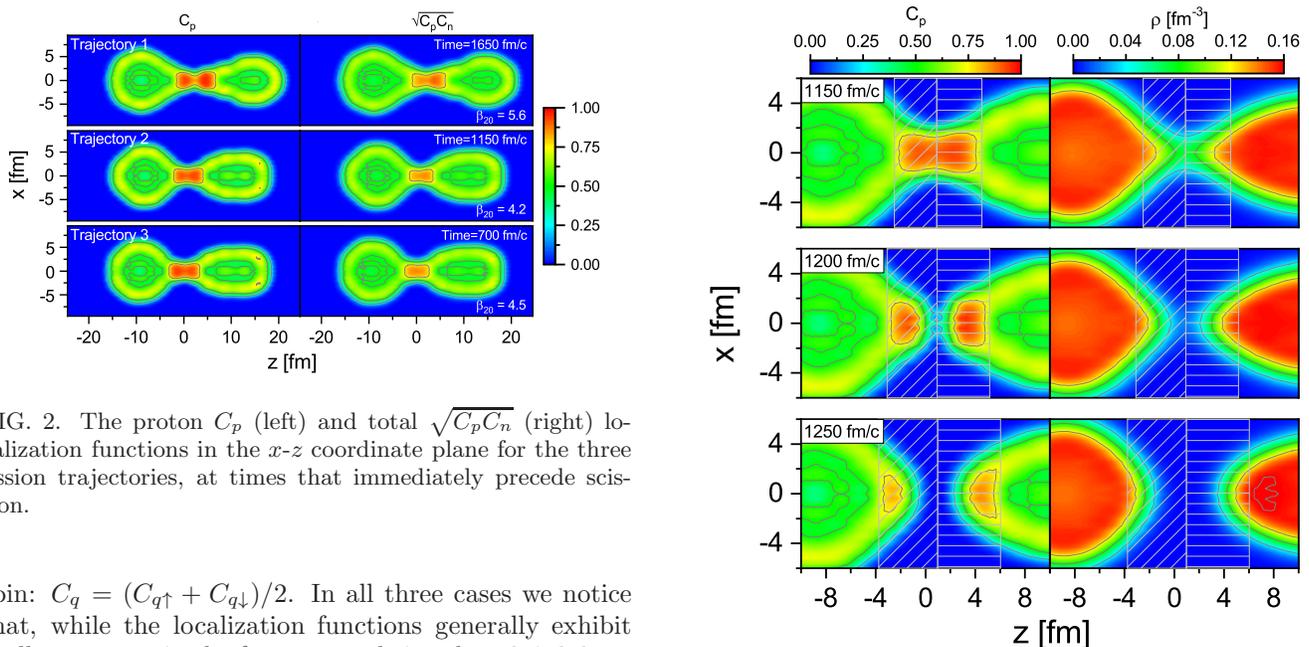}
 \caption{The proton $C_p$ (left) and total $\sqrt{C_p C_n}$ (right) localization functions in the $x$-$z$ coordinate plane for the three fission trajectories, at times that immediately precede scission.}
 \label{fig:localization}
\end{figure}

These findings are consistent with previous studies of fission dynamics but a surprising result is obtained when, instead of the isodensities at prescission times, one considers the corresponding nucleon localization functions \cite{becke90,reinhard11}:
\begin{equation} \label{nlf}
C_{q\sigma}(\vec{r})=\left[1+\left(\frac{\tau_{q\sigma}\rho_{q\sigma}-{1\over 4}|\vec{\nabla}\rho_{q\sigma}|^2-\vec{j}^2_{q\sigma}}{\rho_{q\sigma}\tau^\mathrm{TF}_{q\sigma}}\right)^2\right]^{-1} ,
\end{equation}
for the spin $\sigma$ ($\uparrow$ or $\downarrow$)  and isospin $q$ ($n$ or $p$) quantum
numbers. $\rho_{q\sigma}$, $\tau_{q\sigma}$, $\vec{j}_{q\sigma}$, and $\vec{\nabla}\rho_{q\sigma}$ denote the nucleon density, kinetic energy density, current density, and density gradient, respectively.
$\tau^\mathrm{TF}_{q\sigma}={3\over 5}(6\pi^2)^{2/3}\rho_{q\sigma}^{5/3}$ is the Thomas-Fermi kinetic energy density.

For homogeneous nuclear matter $\tau = \tau^\mathrm{TF}_{q\sigma}$, the second and third term in the numerator vanish,
and $C_{q\sigma} = 1/2$. In the other limit $C_{q\sigma} (\vec{r}) \approx 1$
indicates that the probability of finding two nucleons with the same spin and isospin at the same point $\vec{r}$ is very small.
This is the case for the $\alpha$-cluster of four particles: $p \uparrow$,  $p \downarrow$, $n \uparrow$,
and $n \downarrow$, for which all four nucleon localization functions $C_{q\sigma} \approx 1$. The nucleon localization functions have been used to analyze $\alpha$ cluster structures in light nuclei \cite{reinhard11, ebran17, ren20}, to characterize shell structures of nascent fragments in fissioning nuclei \cite{zhang16,scamps18}, and cluster structures in complex precompound states formed in heavy-ion fusion reactions \cite{Naz17}.

In Fig.~\ref{fig:localization}, we plot the proton $C_p$ (left) and total $\sqrt{C_p C_n}$ (right) localization functions in the $x$-$z$ coordinate plane for the three fission trajectories discussed above, at times that immediately precede scission. Here, the proton and neutron total localization functions are averaged over the spin: $C_{q}=(C_{q\uparrow}+C_{q\downarrow})/2$. In all three cases we notice that, while the localization functions generally exhibit shell structures in the fragments, their values $0.4$--$0.6$ are consistent with homogeneous nuclear matter.
In the neck region, however, values close to 1 are obtained, characteristic for $\alpha$ clusters. It appears as if, at times immediately prior to the scission event, two $\alpha$ particles (or other light clusters, $^{3}$H and $^{6}$He) are present in the neck between the fragments. Actually this is not surprising considering that the nucleon density in the neck region is strongly reduced compared to the nuclear matter saturation density $\approx 0.15$ fm$^{-3}$ (cf. Fig.~\ref{fig:240PES}). It is well known that at subsaturation densities correlations in strongly interacting matter lead to clustering phenomena \cite{zinner13} and, eventually, to a gas phase with nucleons and light clusters.

\begin{figure}[tbh!]
\centering
\includegraphics[width=0.45\textwidth]{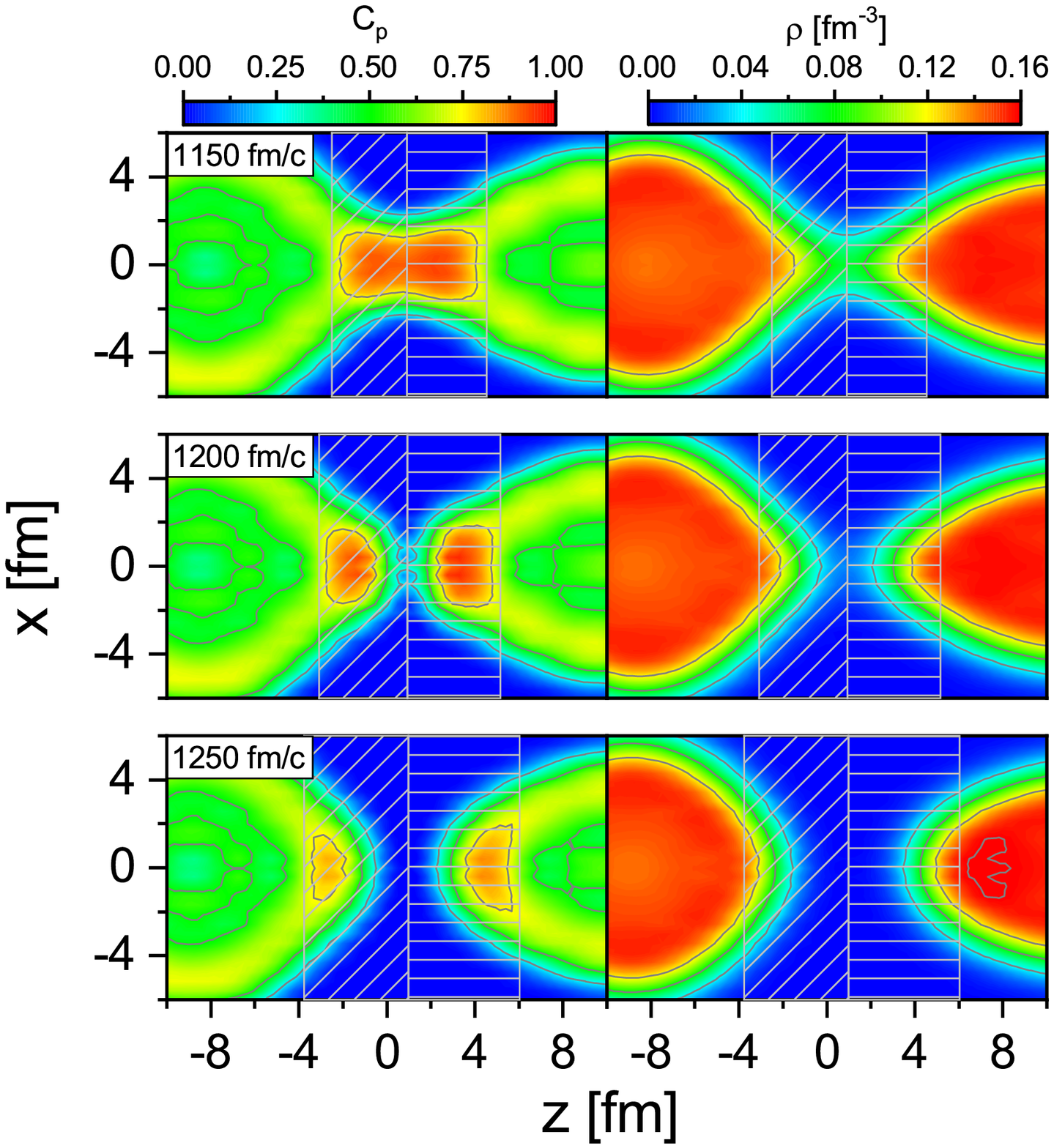}
 \caption{The proton localization function $C_p$ (left) and total density (right), at times: 1150, 1200, and 1250 fm/c, for the fission trajectory 2. Starting from the point of lowest density along the $z$ axis, the shaded areas on the left and on the right denote regions that contain exactly two protons each.}
 \label{fig:neck_localization}
\end{figure}
In principle, a similar result for the localization function could be obtained in static self-consistent calculations with constraints on multipole moments. However, we have found that when the static energy surface is constrained with the values of $\beta_{20}$ and $\beta_{30}$ obtained from the dynamical calculation at the time of scission (1150 fm/c for trajectory number 2), the corresponding proton density and localization function exhibit fragments that are already separated (see the SM~\cite{SM}). The dynamic scission configuration is not reproduced by a static calculation because the latter corresponds to the lowest energy configuration, whereas the former has generally higher energy (the fragments are not cold). Stated differently, the adiabatic assumption on which the static self-consistent mean-field calculation is based, is no longer valid close to scission where nonadiabatic effects are crucial. As illustrated in Ref.~\cite{simenel18}, in the adiabatic approximation the system can advance from a prescission to the postscission configuration in a single step.

When are the light clusters observed in the neck region formed? What is their structure? What is their role in the scission mechanism? In the following we present illustrative results for trajectory number 2 (cf. Fig.~\ref{fig:240PES}), while those for the other two fission trajectories are included in the SM~\cite{SM}. Figure \ref{fig:neck_localization} displays the proton localization function $C_p$ (left) and total nucleon density (right) at three times: immediately preceding scission (1150 fm/c), at the moment when the fragments separate (1200 fm/c), and immediately after (1250 fm/c) when the separated fragments accelerate because of Coulomb repulsion.
For each case, we have determined the point of lowest density along the $z$ axis and, starting from that point, the shaded areas on the left and on the right correspond to the regions that contain exactly two protons. There are, of course, more neutrons in these regions. Before scission, at 1150 fm/c, the calculated number of neutrons in the left shaded area is 4.1 (heavier fragment), and 3.8 on the right (lighter fragment). At the two later times these numbers increase to 4.3 (left) and 3.9 (right). It is interesting to note that, while there is no signature of clustering in the total nucleon densities, the localization function clearly identifies the presence of two $\alpha$-like particles in the neck prior to scission. In fact, the elongation of the neck corresponds to the region that contains, in total, four protons. The two $\alpha$ clusters can also be distinguished at scission (1200 fm/c), while at 1250 fm/c they are already absorbed by the fragments. We have also followed the evolution of the separated fragments at later times, when the absorbed $\alpha$ clusters induce strongly damped oscillations along the fission axis (see the time-evolution movies included in the SM~\cite{SM}).

\begin{figure}[tbh!]
\centering
\includegraphics[width=0.45\textwidth]{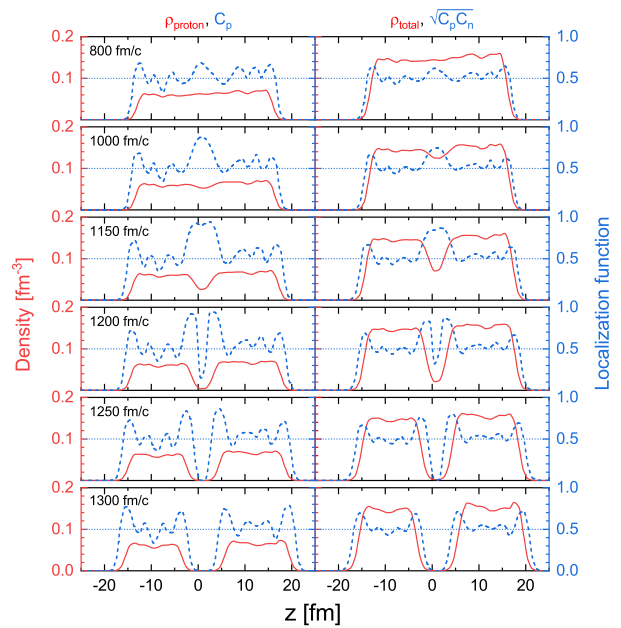}
 \caption{The proton density and localization function (left), and the total density and localization function (right), as functions of the distance along the fission axis, at six characteristic times after the evolution started along fission path 2. The red solid curves denote the nucleon densities, while the localization functions are represented by blue dashed curves.}
 \label{fig:time_evolution}
\end{figure}

To determine the timescale for the formation of $\alpha$ particles in the neck, we follow the evolution of the density and localization functions as the system advances towards fission. In Fig.~\ref{fig:time_evolution} we plot the proton density and localization function (left), and the total ones (right), as functions of the distance along the fission axis, at six characteristic times after the evolution started along fission path 2 (snapshots of the time-evolution movies included in the SM~\cite{SM}).
For the first 1000 fm/c the nucleus evolves in the fission valley and becomes more deformed.
Both the axial elongation and the mass asymmetry increase, but the densities change very little along the $z$ axis, and the localization functions oscillate around the value of 0.5 (dotted horizontal line), characteristic for homogeneous nuclear matter.  At about 1000 fm/c a small dip appears in the central part of the densities and, in the same interval, the localization functions display a marked increase over the homogeneous matter value. At 1150 fm/c, immediately preceding scission, the nucleon density between the fragments is only about half the nuclear matter saturation density, and the proton localization function is close to one, characteristic for the $\alpha$-like cluster. The total localization is somewhat less pronounced because, as noted above, there are almost twice as many neutrons as protons in the neck region. At the moment of scission (1200 fm/c), the two fragments separate but one also identifies two $\alpha$-like clusters between the fragments. The most important result here is that scission appears to occur at the time when the two $\alpha$-like clusters form, and the neck ruptures exactly between the two clusters. This is not difficult to understand because, as the two clusters separate even a small distance, their Coulomb repulsion prevails over the nuclear attraction, and the clusters are forced back toward the fragments. Note that this mechanism is only exposed by the time-evolution of the localization functions, while the density profiles do not exhibit any signature of clustering. Even more clearly, the time-scale of the formation of clusters and the scission event are identified in the time evolution movies included in the SM~\cite{SM}.
Therefore, from the localization functions we deduce that the time scale for the formation of $\alpha$-like clusters in the neck region is $\approx 100 - 200$ fm/c. This process is very rapid compared to the typical time the system evolves along a TDDFT fission trajectory, and even more so in comparison with the fission timescale starting from the compound system in quasiequilibrium \cite{hinde92}.

Essentially the same results are obtained along the two other fission trajectories (included in the SM~\cite{SM}, together with the corresponding time-evolution movies) and, thus, it appears that the formation of light clusters in the neck region and the resulting scission mechanism presents a general attribute of fission dynamics.
It is also possible that one of the $\alpha$-like clusters is not absorbed by the respective fragment and, therefore, it will materialize as the third fragment in the process of ternary fission.
In the SM~\cite{SM} we also include examples of fission trajectories which start from nonaxial shapes.

In conclusion, nuclear TDDFT has been employed to study the dynamics of neck formation and rupture in the process of induced nuclear fission. By following mass-asymmetric fission trajectories in $^{240}$Pu, we have shown that the timescale of neck formation coincides with the assembly of two $\alpha$-like clusters ($\approx 100 - 200$ fm/c).
The low-density region between the nascent fragments provides conditions for dynamical synthesis of $^{4}$He and other light clusters.
The neck ruptures at a point exactly between the two $\alpha$-like clusters, which separate because of the Coulomb repulsion and are eventually absorbed by the two emerging fragments.
In general, elongated configurations at scission facilitate the emergence of a low-density neck region in which clusters of nucleons are formed. Very compact scission configurations such as, for instance, the symmetric scission of $^{264}$Fm, will tend to suppress the formation of clusters between nascent fragments. The newly proposed scission mechanism opens exciting possibilities for a microscopic study of ternary fission.

\begin{acknowledgments}
This work has been supported in part by the High-end Foreign Experts Plan of China,
National Key R\&D Program of China (Contracts No. 2018YFA0404400 and No. 2017YFE0116700),
the National Natural Science Foundation of China (Grants No. 12070131001, No. 11875075, No. 11935003, No. 11975031, and No. 12141501),
the China Postdoctoral Science Foundation under Grant No. 2020M670013,
the High-performance Computing Platform of Peking University,
the QuantiXLie Centre of Excellence, a project cofinanced by the Croatian Government and European Union through the European Regional Development Fund the Competitiveness and Cohesion Operational Programme (KK.01.1.1.01.0004),
 and the Croatian Science Foundation under the project Uncertainty quantification within the nuclear energy density framework (IP-2018-01-5987).
J. Z. acknowledges support by the National Natural Science Foundation of China under Grants No. 12005107 and No. 11790325.
\end{acknowledgments}

\end{document}